\documentclass[
aps,%
12pt,%
final,%
notitlepage,%
oneside,%
onecolumn,%
nobibnotes,%
nofootinbib,%
superscriptaddress,%
showpacs,%
centertags]%
{revtex4}

\usepackage{amsfonts}

\def\'#1{{\accent19\ifx #1i \i\else #1\fi}}

\def\bea{\begin{eqnarray}}
\def\eea{\end{eqnarray}}

\def\dd{{\mathrm d}}

\def\1{\'{\i}}

\def\jp{J_+}
\def\jm{J_-}
\def\jj{J_3}
\def\otra{b}

 \def\kk{K}

\def\la{\lambda}
 \def\te{\theta}
 \def\d{{\rm d}}
\def\rr{\rho}

\def\SW{\rm I}
 \def\Stc{\rm MS}

\def\pot{{\cal U}}

\def\k{\kappa}
\def\>#1{{\mathbf#1}}

\begin{document}
\phantom{1}\bigskip
\title{Superintegrability on $sl(2)$-coalgebra spaces\footnote{Based on the contribution presented at the ``XII International Conference on Symmetry Methods in Physics",   Yerevan (Armenia),  July  2006.\\
To appear in Physics of Atomic Nuclei.}
}

\author{\firstname{\'Angel}~\surname{Ballesteros}}
\email{angelb@ubu.es}
\affiliation{Departamento de F\'isica, Facultad de Ciencias,  Universidad de Burgos,
E-09006, Burgos, Spain.}
\author{\firstname{Francisco J.}~\surname{Herranz}}
\email{fjherranz@ubu.es}
\affiliation{Departamento de F\'isica, Escuela Polit\'ecnica Superior,
Universidad de Burgos, E-09006, Burgos, Spain.}
\author{\firstname{Orlando}~\surname{Ragnisco}}
\email{ragnisco@fis.uniroma3.it}
\affiliation{Dipartimento di Fisica,   Universit\`a di Roma Tre and Instituto
Nazionale di Fisica Nucleare sezione di Roma Tre,  Via Vasca Navale 84, 
I-00146 Roma, Italy}

\begin{abstract}
\baselineskip=16pt

\phantom{1}\bigskip
We review a recently introduced set of $N$-dimensional  quasi-maximally superintegrable Hamiltonian
systems describing geodesic motions, that can be used to generate ``dynamically" a
large family of curved spaces. From an algebraic viewpoint,  such spaces are obtained through kinetic
energy Hamiltonians defined on either the $sl(2)$  Poisson coalgebra or a
quantum deformation of it. Certain potentials on these spaces and endowed with
the same underlying coalgebra symmetry have been also introduced in such a way
that the superintegrability properties of the full system are preserved.
Several new
$N=2$ examples of this construction are explicitly given, and specific Hamiltonians leading to
spaces of non-constant curvature are emphasized.

\baselineskip=16pt

\end{abstract}
\maketitle

 \baselineskip=16pt

\newpage

\section{Introduction}

Two infinite families of $N$-dimensional ($N$D) quasi-maximally
superintegrable Hamiltonians endowed with a set of $(2N-3)$ integrals
of the motion have been recently introduced in
\cite{BHletter,BHSIGMA,plb,jpa2D, Checz}. In the first family, the
superintegrability properties of all these Hamiltonians are shown to be a
consequence of a hidden 
$sl(2)$ Poisson coalgebra symmetry \cite{BHletter}. The second family  is just a
$q$-deformation of the former (see \cite{BHSIGMA,plb,jpa2D, Checz} and references
therein), and the deformed coalgebra  symmetry is given by $sl_z(2)$ ($q={\rm e}^z$),
the Poisson analogue of the non-standard quantum deformation of $sl(2)$ \cite{Ohn}. As
a concrete application of these general results, some of these  Hamiltonians can
be shown to 
 generate superintegrable geodesic
 motions on certain curved manifolds (see \cite{BHletter,plb, Checz}) . In this contribution, we briefly  review this
approach and  we provide new 2D explicit examples of such $sl(2)$-coalgebra
spaces.
 
\section{$sl(2)$-coalgebra spaces}

We recall that an
 $N$D completely integrable Hamiltonian
$H^{(N)}$ is called {\it maximally superintegrable} (MS) if there exists
a set of
$(2N-2)$ globally defined  functionally independent constants of the motion that
Poisson-commute with $H^{(N)}$. Among them, at least, two
different subsets of
$(N-1)$ constants in involution can be found.
In the same way, 
a system will be called  {\it
quasi-maximally superintegrable} (QMS) if there are  
$(2N-3)$ independent integrals with the abovementioned properties. 

Let us now consider the $sl(2)$ Poisson coalgebra
generated by the following Lie--Poisson brackets and comultiplication map: 
\begin{equation}
 \{J_3,J_+\}=2 J_+     ,\qquad  
\{J_3,J_-\}=-2 J_- ,\qquad   
\{J_-,J_+\}=4 J_3  ,   
\label{ba}
\end{equation}
\begin{equation}
\begin{array}{l}
\Delta(J_l)=  J_l \otimes 1+ 1\otimes J_l ,\qquad l=+,-,3.
\end{array}
\label{bb}
\end{equation}
The Casimir function for $sl(2)$ reads
\begin{equation} 
{\cal C}=  J_- J_+ -J_3^2  . 
\label{bc}
\end{equation} 

The following result holds~\cite{BHletter}:

 \noindent
{\bf Theorem 1.} 
{Let $\{\>q,\>p \}=\{(q_1,\dots,q_N),(p_1,\dots,p_N)\}$ be $N$ pairs of
canonical variables. The $N$D Hamiltonian
\begin{equation}
{H}^{(N)}= {\cal H}\left(J_-,J_+,J_3\right),
\label{hgen}
\end{equation}
with ${\cal H}$ any smooth function and
\begin{equation}
 J_-=\sum_{i=1}^N q_i^2\equiv \>q^2 ,\quad\    J_+=
    \sum_{i=1}^N \left(  p_i^2+\frac{\otra_i}{ q_i^2} \right)
\equiv \>p^2 +  \sum_{i=1}^N  \frac{\otra_i}{ q_i^2}  ,\quad\ J_3=
  \sum_{i=1}^N  q_i p_i\equiv \>q\cdot\>p  ,
\label{qp}
\end{equation}
where $b_i$ are arbitrary real parameters, is a
QMS system. The   $(2N-3)$ functionally
independent and ``universal" integrals of the motion are explicitly given
by
\bea
&& C^{(m)}= \sum_{1\leq i<j}^m \left\{ ({q_i}{p_j} -
{q_j}{p_i})^2 + \left(
\otra_i\frac{q_j^2}{q_i^2}+\otra_j\frac{q_i^2}{q_j^2}\right)\right\}
+\sum_{i=1}^m \otra_i , \nonumber\\
&& C_{(m)}= \sum_{N-m+1\leq i<j}^N \left\{ ({q_i}{p_j} -
{q_j}{p_i})^2 + \left(
\otra_i\frac{q_j^2}{q_i^2}+\otra_j\frac{q_i^2}{q_j^2}\right)\right\}
+\sum_{i=N-m+1}^N \otra_i  ,
\label{cinfm}
\eea
where $m=2,\dots, N$ and $C^{(N)}=C_{(N)}$.  Moreover, the sets of $N$ functions
$\{H^{(N)},C^{(m)}\}$ and $\{H^{(N)},C_{(m)}\}$ $(m=2,\dots, N)$ are in
involution. 
}
\medskip

The proof of this result is based on the fact that, for any
choice of the function
${\cal H}$, the Hamiltonian
$H^{(N)}$ has an $sl(2)$ Poisson coalgebra symmetry (see \cite{BR,Deform}).
Notice that for arbitrary  
$N$ there is a single constant of the motion left  in order to assure   maximal
superintegrability. In case that an additional
integral does exist, the latter does not come from  the coalgebra symmetry and has to
be found directly. Finally, note that in the case $N=2$ quasi-maximal
superintegrability is just equivalent to integrability (the coalgebra symetry provides
only $(2N-3)=1$ integral of the motion). However, these coalgebra systems can, by
construction, be generalized to arbitrary dimension.

Let us now give some explicit examples of the QMS  spaces coming from this
construction as particular geodesic motion Hamiltonians.

\subsection{$sl(2)$-coalgebra spaces with constant curvature}

Let us consider the symplectic realization (\ref{qp})   with $\otra_i=0, \forall i$.
The  kinetic energy  ${\cal T}$ of a particle  on the $N$D  Euclidean space ${{\bf
E}^N}$ can be directly interpreted as the generator
$J_+$. Therefore, the ${{\bf E}^N}$ space can be thought  of as the manifold with
geodesic motions given by
\begin{equation}
 {\cal H} = {\cal T}=\frac 12 \jp =\frac 12\, {\>p}^2 .
 \label{bbff}
\end{equation}

Moreover, the kinetic energy 
  on $N$D Riemannian spaces with constant curvature $\kappa$ can be expressed
in Hamiltonian form as a function of the $sl(2)$ generators  in two different
ways
(see \cite{BHletter} for a detailed geometrical interpretation of this result):
\begin{equation}
\begin{array}{l}
\displaystyle{ {\cal H}^{\rm P}={\cal T}^{\rm P}=\frac{1}{2}\left( 1+\k
J_-\right)^2 J_+=
\frac{1}{2}\left( 1+\k \>q^2\right)^2 \>p^2} ,\\[8pt]
\displaystyle{ {\cal H}^{\rm B}={\cal T}^{\rm B}=\frac{1}{2}\left( 1+\k
J_-\right)\left(  J_+ +\k J_3^2\right)=
\frac{1}{2}(1+\k \>q^2)\left( \>p^2+\k (\>q\cdot \>p)^2 \right) }.
\end{array}
\label{dd}
\end{equation}

The first one ${\cal H}^{\rm P}$ is just the kinetic energy for a
free particle  on  the spherical ${\bf S}^N$  
($\k>0$) and hyperbolic
${\bf H}^N$    ($\k<0$) spaces  in terms of Poincar\'e
coordinates $\>q$ (coming from a stereographic projection in
${\mathbb R^{N+1}}$) and their associated canonical momenta $\>p$. The second one
${\cal H}^{\rm B}$ corresponds to Beltrami coordinates  and momenta (central
projection). In the framework here presented, both Hamiltonians can immediately be 
interpreted as deformations (in terms of the curvature parameter $\k$) of the flat
Euclidean motion given by $\k=0$. By construction, and for any dimension, both 
geodesic motions are QMS ones since they Poisson-commute with the integrals
(\ref{cinfm}).

\subsection{$sl(2)$-coalgebra spaces with non-constant curvature}

Note that, in principle, any homogeneous quadratic  function of the canonical momenta
can provide an admissible   geodesic motion.  In particular, we can consider
the $sl(2)$ Hamiltonian (with $\otra_i=0, \forall i$)
\begin{equation}
 {\cal H}  = {\cal T}=\frac 12 \, f(\jm)\, \jp =\frac 12\, f({\>q}^2)\, {\>p}^2 ,
 \label{var}
\end{equation}
with $f$ an arbitrary smooth function, and we can derive from it the  corresponding
kinetic energy Lagrangian. Hence the QMS geodesic motion is defined on a
Riemannian manifold whose metric is given by
\begin{equation}
\dd s^2=\frac 2{f({\>q}^2)} \, \dd{\bf q}^2 .
\label{ds2}
\end{equation}

In the $N=2$ case we can easily compute the corresponding Gaussian
curvature $K$ of the space; namely
\begin{equation}
 K = \frac{1}{f({\>q}^2)}\left\{
 -{\>q}^2\,f'({\>q}^2)+ f({\>q}^2)\left[  f'({\>q}^2) + {\>q}^2\,f''({\>q}^2)
 \right]
 \right\},
\end{equation}
where $f'$ and $f''$ are the derivatives with respect to the variable $\>q^2=(q_1^2+q_2^2)$.
Therefore, we have obtained an infinite family of $N=2$ spaces with, in general,
non-constant curvature depending on the ``radial" coordinate ${\>q}^2$
(see~\cite{darboux1,darboux2,darbouxresto} for the study of 2D and 3D superintegrable
systems on spaces with non-constant curvature). Obviously, the constant curvature spaces
given   in terms of Poincar\'e coordinates (\ref{dd})  are just   particular cases of
this construction with $f(\jm)=\left( 1+\k J_-\right)^2/2$, that is, $K=\kappa$.

Another
remarkable
$sl(2)$-coalgebra space, contained in (\ref{var}), is obtained by setting
\begin{equation}
 {\cal H}  = {\cal T}=\frac 12 \, \frac{\jp}{\alpha + \jm} =\frac 12\, 
\frac{{\>p}^2}{\alpha + {\>q}^2}.
 \label{darbouxIII}
\end{equation}
The corresponding space is just the so-called Darboux space of type III  \cite{darboux1},
whose $N=2$ non-constant Gaussian curvature reads
\begin{equation}
 K = -\frac{\alpha}{(\alpha+q_1^2+q_2^2)^3}.
\end{equation}
A  MS (intrinsic) Smorodinsky--Winternitz system~\cite{fris,evans2} on the
$N$D generalization of this Darboux space has been recently obtained in
\cite{enciso}.

We stress that the   Hamiltonian (\ref{var}) does not exhaust all the possibilities
for free motion, since $J_3^2= ( \>q\cdot\>p)^2$ is also a quadratic homogeneous
function in the momenta. Therefore, we could consider  more complicated kinetic energy
terms including
$J_3^2$. A particular  choice of this type would be the one
given by the Beltrami  kinetic energy presented in (\ref{dd}), for which
the metric of the space reads
\begin{equation}
\dd s^2= \frac{(1+\k\>q^2)\dd\>q^2-\k
(\>q\cdot \dd\>q)^2}{(1+\k\>q^2)^2} ,
\end{equation}
and all its sectional curvatures are constant and equal to $\k$.
 Another choice is
given by the Hamiltonian
\begin{equation}
 {\cal H}=\frac{1}{2}\, J_+  + \alpha J_3^2,
\label{conj3}
\end{equation}
where $\alpha$ is a real parameter. For $N=2$ this gives rise   to a space with a
metric
\begin{equation}
\dd s^2=  \frac {2}{1+2\,\alpha\,(q_1^2+q_2^2)}\left\{
 {(1+2\,\alpha\,q_2^2)}\,\d q_1^2   +
  {(1+2\,\alpha\,q_1^2)}\, \d q_2^2 
  -4\,\alpha\, {q_1\,q_2}\,\d q_1\,\d q_2\right\},
\end{equation}
  also endowed with a constant Gaussian curvature
$K=-\alpha$. In contrast to the latter case, if one considers the free Hamiltonian
\begin{equation}
{\cal H}=\frac{1}{2}J_+ +\alpha\,J_-\, J_3^2=
\frac{1}{2}\>p^2+\alpha\, \>q^2\, (\>q\cdot \>p)^2 ,
\label{conj3b}
\end{equation}
one finds a space whose metric for $N=2$ reads
\begin{equation}
\dd s^2=   \frac {2(1+2\,\alpha\,q_2^2\,(q_1^2+q_2^2))}{1+2\,\alpha\,(q_1^2+q_2^2)^2}\,\d q_1^2   +
  \frac {2(1+2\,\alpha\,q_1^2\,(q_1^2+q_2^2))}{1+2\,\alpha\,(q_1^2+q_2^2)^2}\, \d q_2^2 
  -8\,\alpha\,\frac {q_1\,q_2\,(q_1^2+q_2^2)}{1+2\,\alpha\,(q_1^2+q_2^2)^2}\,\d q_1\,\d q_2,
\end{equation}
and whose nonconstant Gaussian curvature is found to be  
\begin{equation}
 K = -2\,\alpha\,(q_1^2+q_2^2).
\end{equation}
In any case, we stress that the geometric interpretation of  the canonical variables
$(\>q,\>p)$ can be completely different for each $sl(2)$-coalgebra space.

\subsection{QMS potentials}

The underlying coalgebra symmetry of this construction is also  helpful in order to
define QMS potentials ${\cal V}$ on $sl(2)$-coalgebra spaces. This can be achieved, in
general, by adding some suitable functions depending on
$J_-$ to the kinetic energy term and by considering arbitrary centrifugal  terms that
come from symplectic realizations of the $J_+$ generator with
constants $\otra_i$'s that are different from zero. Such a  Hamiltonian would be given
by
\begin{equation}
{\cal H}={\cal T}(J_+,J_-,J_3) + {\cal V}(J_-).
\label{potential}
\end{equation}
In the constant curvature case, the Hamiltonians that we would obtain in this way are
the curved counterpart of the Euclidean systems for different values of the
sectional curvature
$\k$, that lead to QMS potentials on the spaces ${\bf S}^N$   ($\k>0$),
${\bf H}^N$    ($\k<0$), and ${\bf E}^N$   
 ($\k=0$). A detailed description for such potentials can be found in \cite{BHletter}.  The
same scheme can be applied to the non-constant  curvature spaces given in Section II.B. In all the
cases, the fact that the Hamiltonian (\ref{potential}) is defined on the ``abstract"
$sl(2)$-coalgebra generators ensures the existence of the  set of $(2N-3)$ 
``universal" integrals given in theorem 1, whatever the functions ${\cal T}$ and ${\cal
V}$ be.

\section{$sl_z(2)$-coalgebra spaces}

We recall that the non-standard
$sl_z(2)$ Poisson coalgebra is given by the following
deformed Poisson brackets and   coproduct~\cite{Deform}:
\begin{equation}
\{\jj,\jp\}=2 \jp \cosh z\jm  ,\qquad
 \{\jj,\jm\}=-2\,\frac {\sinh z\jm}{z} ,\qquad
 \{\jm,\jp\}=4 \jj ,
\label{baa}
\end{equation}
\begin{equation}
  \Delta_z(\jm)=  \jm \otimes 1+
1 ,\qquad  \Delta_z(J_l)=J_l \otimes {\rm e}^{z \jm} + {\rm e}^{-z \jm} \otimes
J_l ,\qquad l=+,3.
\label{ggcc}
\end{equation}
The Casimir function for $sl_z(2)$ reads
\begin{equation}
\qquad\qquad{\cal C}_z=  \frac {\sinh z\jm}{z}\, \jp -\jj^2 .
\label{gc}
\end{equation}

The construction presented in the previous Section  can be
generalized to the case of this quantum deformation of  $sl(2)$, and
the associated spaces will be, in general, of non-constant  curvature. Explicitly,  we have
the following general result~\cite{Deform,BHSIGMA}:

 \noindent
{\bf Theorem 2.} 
Let $\{\>q,\>p \}=\{(q_1,\dots,q_N),(p_1,\dots,p_N)\}$ be $N$ pairs of
canonical variables. The $N$D Hamiltonian
\begin{equation}
{H}^{(N)}_z= {\cal H}_z\left(\jm^{(N)},\jp^{(N)},\jj^{(N)}\right),
\label{hgenb}
\end{equation}
where ${\cal H}_z$ is any smooth function and
\bea
&& 
\jm^{(N)}= \sum_{i=1}^N q_i^2 \equiv \>q^2 ,\qquad
\jj^{(N)}=\sum_{i=1}^N
\frac {\sinh z q_i^2}{z q_i^2} \, q_ip_i \, {\rm e}^{z \kk_i^{(N)}(q^2) } 
\equiv (\>q\cdot\>p)_z ,\nonumber \\ 
&& \jp^{(N)}=\sum_{i=1}^N
\left( \frac {\sinh z q_i^2}{z q_i^2} \, p_i^2  +\frac{z \otra_i}{\sinh z
q_i^2} \right)  {\rm e}^{z \kk_i^{(N)}(q^2) }\equiv \tilde{\>p}_z^2  ,
\label{zsymp}
\eea
where \begin{equation}
  \kk_i^{(N)}(q^2)=  - \sum_{k=1}^{i-1}  q^2_k+ 
\sum_{l=i+1}^N   q^2_l ,
\label{kki}
\end{equation}
is 
QMS for any choice of the function ${\cal H}$ 
and for arbitrary real parameters $b_i$.

We remark that the explicit expressions for the
$(2N-3)$ functionally independent and ``universal" integrals of the motion
   can be found in \cite{BHSIGMA}. 

Let us explicitly write the
$2$-particle symplectic realization of $sl_z(2)$ (\ref{zsymp}):
\bea
&& \jm^{(2)}=q_1^2+q_2^2  , \qquad \jj^{(2)}=
 \frac {\sinh z q_1^2}{z q_1^2 } \, {\rm e}^{z q_2^2}   q_1 p_1  +
 \frac {\sinh z q_2^2}{z q_2^2 }\, {\rm e}^{-z q_1^2}    q_2 p_2  ,\nonumber\\
&& \jp^{(2)}=
  \frac {\sinh z q_1^2}{z q_1^2}\,  {\rm e}^{z q_2^2}   p_1^2   +
 \frac {\sinh z q_2^2}{z q_2^2} \,  {\rm e}^{-z q_1^2}   p_2^2 +\frac{z
\otra_1}{\sinh z q_1^2}\,{\rm e}^{z q_2^2} +
\frac{z \otra_2}{\sinh z
q_2^2} \,  {\rm e}^{-z q_1^2} .
\label{sympz2} 
\eea
In this case there is a single  constant of the motion:
\begin{equation} 
C_z^{(2)}= \frac
{\sinh z\jm^{(2)}}{z}\, \jp^{(2)} - \left(\jj^{(2)}\right)^2  . 
\label{constant}
\end{equation} 

Again we are dealing with free motion, thus we will take the symplectic realization with
$\otra_1=\otra_2=0$ in order to avoid centrifugal terms.
In a parallel way to (\ref{var}), we can consider an infinite family of {\em integrable} 
(and quadratic
 in the momenta) free $N=2$ motions with $sl_z(2)$-coalgebra symmetry
through Hamiltonians of the type
\begin{equation}
 {H}^{(2)}_z =\frac 12 \jp^{(2)} f\bigl(z\jm^{(2)}\bigr) ,
 \label{bbfff}
\end{equation}
where $f$ is an arbitrary smooth function such
that 
 $\displaystyle{\lim_{z\to 0}f\bigl(z\jm^{(2)}\bigr)=1}$, that is,
$\lim_{z\to 0}{H}^{(2)}_z=\frac 12 (p_1^2+p_2^2)$.  We shall explore in
the sequel some specific choices for $f$, and we shall analyse the spaces generated by
them.


\subsection{An $sl_z(2)$-coalgebra space with non-constant  curvature}

Of course, the
 simplest choice will be just to set $f\equiv 1$ \cite{plb}:
\begin{equation}
{\cal H}^{\rm \SW}_z=\frac 12
 \jp^{(2)}=\frac12 \left( \frac {\sinh z
 q_1^2}{z q_1^2} \, {\rm e}^{z q_2^2}  p_1^2   +
 \frac {\sinh z q_2^2}{z q_2^2} \, {\rm e}^{-z q_1^2}  p_2^2  \right)   .
 \label{bg}
\end{equation}
Hence the {kinetic energy} ${\cal T}^{\rm
\SW}_z(q_i,p_i)$ coming from
 ${\cal H}^{\rm \SW}_z$  is 
\begin{equation}
 {\cal T}^{\rm \SW}_z(q_i,\dot q_i)=\frac 12 \left(\frac
 {z q_1^2}{\sinh z q_1^2} \, {\rm e}^{-z q_2^2} \dot q_1^2   +
 \frac {z q_2^2}{\sinh z q_2^2} \, {\rm e}^{z q_1^2} \dot q_2^2  
 \right) ,
 \label{ca}
\end{equation}
 and defines a geodesic flow on a 2D
 Riemannian space with signature  
 diag$(+,+)$ and metric given by:
\begin{equation}
 \d s_I^2=\frac {2z q_1^2}{\sinh z
 q_1^2} \, {\rm e}^{-z q_2^2} \,\d q_1^2   +
  \frac {2 z q_2^2}{\sinh z q_2^2} \, {\rm e}^{z q_1^2}\, \d q_2^2  .
 \label{cc}
\end{equation}
 
The Gaussian curvature $K$ for this space  turns out to be {nonconstant and {\em
negative}}:
\begin{equation}
 K(q_1,q_2;z)=-   z \sinh\left(z(q_1^2+q_2^2\right))=-   z \sinh\left(z\, \>q^2 \right) .
 \label{cd}
\end{equation}
Thus, the 
 underlying 2D space is of {hyperbolic  type} and  
 endowed with  a {``radial" symmetry}. 

We consider the following transformation that includes a new
parameter
$\la_2\ne 0$: 
\begin{equation}
 \cosh(\la_1 \rr)=\exp\left\{z(q_1^2+q_2^2)\right\} ,\qquad
  \sin^2(\la_2 \te)=\frac{\exp\left\{2z q_1^2
 \right\}-1}{\exp\left\{2z(q_1^2+q_2^2)\right\}-1},
\label{change}
\end{equation}
 where $z=\lambda_1^2$  and both $\la_1,\la_2$ can take either a real or a
 pure imaginary value. 
Note that the zero-deformation limit
$z\to 0$  is in fact the flat   contraction $K \to 0$. Under   this limit 
\begin{equation}
\rho \to
2(q_1^2+q_2^2), \qquad \sin ^2 (\lambda_2 \theta) \to \frac{q_1^2}{q_1^2 +
q_2^2}.
\end{equation}
Thus $\rr$ can be interpreted as  a {radial coordinate} and $\te$
 is a {either circular ($\la_2$ real) or  hyperbolic angle}  ($\la_2$ imaginary). Notice
that in the latter case, say $\la_2={\rm i}$, the coordinate $q_1$ is imaginary 
and can be written as $q_1={\rm i} \tilde q_1$ where $\tilde q_1$ is a real coordinate;
then $\rho \to 2(q_2^2-\tilde q_1^2)$ which corresponds to a relativistic radial
distance. Therefore the introduction of the additional parameter $\la_2$ will allow us to
obtain Lorentzian metrics.

 In
this new coordinates, the metric (\ref{cc}) reads 
\begin{equation}
 \d s_I^2=\frac {1}{\cosh(\la_1 \rr)}
 \left( \d \rr^2  +\la_2^2\,\frac{\sinh^2(\la_1 \rr)}{\la_1^2} \, \d
 \te^2  \right) =\frac {1}{\cosh(\la_1 \rr)}\,\d s_0^2.
 \label{cg}
\end{equation}
where $\d s_0^2$ is just the metric of the 2D Cayley--Klein spaces
 in terms of geodesic polar coordinates~\cite{ramon,Conf}
  provided that we identify $z=\la_1^2\equiv -\kappa_1$ and
 $\la_2^2\equiv \kappa_2$; hence $\la_2$  determines
the signature of the metric. The {Gaussian
curvature} turns out to be
\begin{equation}
 K(\rr)=-\frac 12 \la_1^2 \,\frac{\sinh^2(\la_1 \rr)}{\cosh(\la_1
 \rr)} .
 \label{cj}
\end{equation}

In this way we  find the {following   spaces}:

\begin{itemize}

 \item[$\bullet$]  When  $\la_2$ is real, we get a  2D {deformed
 sphere} ${\bf S}^2_z$ $(z<0)$,    and  a {deformed
 hyperbolic or  Lobachewski space} ${\bf H}^2_z$ $(z>0)$.

 \item[$\bullet$]  When  $\la_2$ is imaginary, we obtain a {deformation
 of   the (1+1)D   anti-de Sitter spacetime} ${\bf AdS}_z^{1+1}$ $(z<0)$ and
of the   {de Sitter one} ${\bf dS}_z^{1+1}$
 $(z>0)$.

 \item[$\bullet$] In the nondeformed case $z\to 0$, we recover the Euclidean
 space
 ${\bf E}^2$ 
 ($\la_2$ real) and  Minkowskian spacetime
 ${\bf M}^{1+1}$ ($\la_2$ imaginary).

 \end{itemize}

In the new variables the kinetic energy  (\ref{ca}) is transformed into
\begin{equation}
 {\cal T}^{\rm \SW}_z(\rr,\te;\dot \rr,\dot \te)=\frac
 {1}{2\cosh(\la_1 \rr)}
 \left(\dot \rr^2  +\la_2^2\,\frac{\sinh^2(\la_1 \rr)}{\la_1^2} \,
 \dot \te^2 
 \right) ,
\end{equation}
and  the 
free motion Hamiltonian (\ref{bg}) is written as
\begin{equation}
 \widetilde{H}^{\SW}_z=\frac 12 \cosh(\la_1
 \rr)\left(p_\rr^2 +\frac{\la_1^2}{\la_2^2\sinh^2(\la_1 \rr)} \, 
 p_\te^2\right),
 \label{dda}
\end{equation}
where $\widetilde{H}^{\SW}_z =2 {\cal H}^{\SW}_z$. The  {unique} constant of the
motion ${ C}^{(2)}_z$ (\ref{constant}) is simply given by 
\begin{equation}
  \widetilde{C}_z=p_\te^2 ,
\end{equation}
provided that $\widetilde{C}_z= 4\la_2^2  { C}^{(2)}_z$, and through
the usual   radial-symmetry reduction we find
\begin{equation}
 \widetilde{H}^{\SW}_z =\frac 12 \cosh(\la_1
 \rr)\, p_\rr^2  +\frac{\la_1^2 \cosh(\la_1
 \rr)}{2\la_2^2\sinh^2(\la_1 \rr)} \, \widetilde{C}_z.
 \label{dede}
\end{equation}


\subsection{$sl_z(2)$-coalgebra spaces with constant curvature}
 
If we consider the function $f\bigl(z\jm^{(2)}\bigr)={\rm e}^{ z \jm^{(2)}}$ we
obtain  a MS 
$sl_z(2)$-coalgebra  Hamiltonian given by  
\begin{equation}
  {\cal H}^{\rm MS}_z =\frac 12  \jp^{(2)} {\rm e}^{ z \jm^{(2)}}=\frac12
\left(
\frac {\sinh z
 q_1^2}{z q_1^2} \, {\rm e}^{z q_1^2}{\rm e}^{2z q_2^2}  p_1^2   +
 \frac {\sinh z q_2^2}{z q_2^2} \, {\rm e}^{ z q_2^2}  p_2^2  \right) . 
 \label{bi}
\end{equation}
Its maximal superintegrability  comes from the existence of an additional (and 
functionally independent)
 constant of the motion given by~\cite{Deform}:
\begin{equation}
 {\cal I}_z=\frac {\sinh z
 q_1^2}{2 z q_1^2} \, {\rm e}^{z q_1^2}  p_1^2 .
 \label{bjjzz}
\end{equation}
 
In this case the kinetic energy Lagrangian is given by
\begin{equation}
 {\cal T}^{\rm MS}_z(q_i,\dot q_i)=\frac 12 \left(\frac {z
 q_1^2}{\sinh z q_1^2} \, {\rm e}^{-z q_1^2}{\rm e}^{-2 z q_2^2} \,\dot
 q_1^2   +
 \frac {z q_2^2}{\sinh z q_2^2} \, {\rm e}^{-z q_2^2}\, \dot q_2^2  
 \right)  ,
 \label{ea}
\end{equation}
whose associated {metric} reads 
\begin{equation}
 \d s_{\rm MS}^2=\frac {2z q_1^2}{\sinh z
 q_1^2} \, {\rm e}^{-z q_1^2}{\rm e}^{-2 z q_2^2} \,\d q_1^2   +
  \frac {2 z q_2^2}{\sinh z q_2^2} \, {\rm e}^{-z q_2^2} \, \d q_2^2  
 .
 \label{eecc}
\end{equation}

Surprisingly enough, the computation of the Gaussian
curvature $K$ for $\d s_{\rm MS}^2$ gives that $K=z$. Therefore, we are dealing
with a space of constant curvature which is just  the deformation
parameter $z$. In \cite{plb} it was shown that a certain change of
coordinates (of the type (\ref{change}) and that includes the signature parameter
$\la_2$) transforms the metric (\ref{eecc}) into
 \begin{equation}
 \d s_{\rm MS}^2= 
  \d r^2  +\la_2^2\,\frac{\sin^2(\la_1 r)}{\la_1^2} \, \d \te^2   ,
 \label{eh}
\end{equation}
which exactly coincides with the metric $\d s_0^2$ of the Cayley--Klein spaces written
in  geodesic polar coordinates
 $(r,\te)$   provided that now $z=\la_1^2\equiv \kappa_1$ and
 $\la_2^2\equiv 
 \kappa_2$.
Obviously, after this change the geodesic motion can be reduced to a
``radial" 1D system:
\begin{equation}
 \widetilde{H}^{\Stc}_z=\frac 12 \, p_r^2
 +\frac{\la_1^2}{2\la_2^2\sin^2(\la_1 r)} \, 
  \widetilde{C}_z  ,
 \label{ffnn}
\end{equation}
where $\widetilde{H}^{\Stc}_z=2{\cal H}^{\Stc}_z$ and  $\widetilde{C}_z=p_\te^2$
is again the generalized momentum for the $\te$ coordinate.


It can also be  checked that other choices for the
Hamiltonian yield constant curvature spaces. In fact, let us consider the
generic Hamiltonian (\ref{bbfff}) depending on $f$,  whose  2D Gaussian curvature can
be expressed in terms of the function
$f(x)$ as
 \begin{equation}
K(x)=z\left(f^\prime(x)\cosh x  +\left(
 f^{\prime\prime}(x)-f(x)-\frac{{f^\prime}^2(x)}{f(x)}
 \right) \sinh x
 \right) ,
 \end{equation}
 where $x\equiv z\jm^{(2)}=z(q_1^2+q_2^2)$. In
general, we obtain $sl_z(2)$-coalgebra spaces with non-constant  curvature. 
In order to characterize the constant curvature cases~\cite{BHSIGMA}, we define $g:=
f^\prime/f$ and we get the nonlinear differential equation
$$
K/z = f^\prime \cosh x + \left(f^{\prime \prime} -f - (f^\prime)^2/f \right) \sinh
x  =  f\left( g \cosh x + (g^\prime -1)\sinh x \right).
$$
If we now  require $K$ to be a constant we obtain the equation
$$
K^\prime=0 \equiv  2y\cosh x + y^\prime \sinh x = 0 ,
\qquad \mbox{where}\quad y:=2g^\prime + g^2
-1 .
$$
The solution for this equation yields 
$${{y=\frac{A}{\sinh^2 x}}},
$$
where $A$ is a constant, and 
solving for ${{g}}$, we  find for ${{F:=f^{\frac{1}{2}}}}$ that
$$
F^{\prime \prime}= \frac{1}{4} \left( 1+ {\frac{A}{\sinh ^2x}}\right)F ,
$$
whose  general solution is (${{A:=\lambda(\lambda-1)}}$):
\begin{equation}
{{F= (\sinh x)^\lambda\left\{ C_1 \bigl(\sinh (x/2)\bigr)^{(1-2\lambda)} + C_2\bigl(
\cosh (x/2)\bigr)^{(1-2\lambda)}\right\} }},
\end{equation}
where $C_1$ and $C_2$ are two integration constants.

Now, if we impose that  
$\lim_{x\to 0}{f}=1$ we obtain that only the cases with $A=0$ are
possible, that is, either $\lambda=1$ or $\lambda=0$. Hence the two elementary
solutions   are 
\begin{equation}
{\cal H}_z=\frac 12  \jp^{(2)} {\rm e}^{\pm  z \jm^{(2)}},
\end{equation}
and the Gaussian curvature of their associated 2D spaces is $K=\pm z$.

\subsection{Other $sl_z(2)$-coalgebra spaces}

Many other possibilities for the definition of the free motion   Hamiltonian in terms
of the $sl_z(2)$-coalgebra generators are indeed possible. In general, such choices would
lead to non-constant  curvature spaces with QMS geodesic motions. For instance,  we can
consider the  kinetic energy Hamiltonian given by
\begin{equation}
 {\cal H}=\frac{1}{2}\, J_+  + \alpha J_3^2 .
\label{conj3z}
\end{equation}
A straightforward calculation shows that, in the $N=2$ case, the Gaussian curvature of
the associated space is just 
$$
K=\frac{\alpha}{2} - \frac{3\,\alpha}{2} \cosh\left(2\,z(q_1^2+q_2^2) \right)
-   z \sinh\left(z(q_1^2+q_2^2) \right) .
$$
Obviously, the nondeformed $z\to 0$ limit of this Hamiltonian   is just the one given
by (\ref{conj3}) and its limiting Gaussian curvature is $-\alpha$. Through this example
we see again that the introduction of the quantum deformation leads to an ``algebraic
generation" of nonconstant curvature on the underlying space.


\subsection{QMS potentials}

As we have just commented, we
can also consider
more general $N$D QMS Hamiltonians based on $sl_z(2)$ (\ref{zsymp}) by
considering arbitrary
$b_i$'s (contained in $J_+$) and by adding some functions  depending on $J_-$.  In
particular, we have considered the Hamiltonians (see
\cite{jpa2D} for the 2D construction):
\begin{equation}
{\cal H}_z=\frac 12 \jp \, f (z\jm 
 )+\pot (z\jm )  ,
\label{ahaa}
\end{equation}
where the arbitrary smooth functions $f$ and $\cal U$ are such that 
\begin{equation}
\lim_{z\to 0}\pot(zJ_-)={\cal V}(J_-), \qquad 
\lim_{z\to
0}f(zJ_-)=1.
\end{equation}
This, in turn, means that
\begin{equation}
\lim_{z\to 0} {\cal H}_z= \frac 12 \,\>p^2 + {\cal V}(\>q^2)
+\sum_{i=1}^N\frac{b_i}{2q_i^2},
\end{equation}
recovering the superposition of a central potential  ${\cal V}(J_-)\equiv {\cal
V}(\>q^2)$ with
$N$ centrifugal terms on  ${\> E}^N$~\cite{Evansa}. 

To end with, we stress that within this construction the function $f (z\jm 
 )$ fixes the type of 
{curved background}, which is characterized
by the metric  $\d s^2/f (z\>q^2 )$
(where $\d s^2$ is the non-constant  curvature metric associated to ${\cal
H}_z=\frac 12 \jp$). Among this infinite family of spaces, the two special cases with $f (z\jm 
 )={\rm e}^{\pm z J_-}$ give rise to Riemannian spaces of constant sectional
curvatures, all of them equal to $\pm z$. 

Particular choices of the potential function for  different $sl_z(2)$-coalgebra spaces
have been proposed and analysed from a geometrical viewpoint in
\cite{jpa2D,BHSIGMA,enciso}. Among them, the non-constant  curvature analogues of the
Smorodinsky--Winternitz and generalized Kepler--Coulomb potentials have been proposed, and many of
the well-known results concerning both potentials for the constant curvature
spaces~\cite{VulpiLett,CRMVulpi,RS,PogosClass1,PogosClass2,KalninsH2,Higgs,Leemon,
Schrodingerdual,Schrodingerdualb,kiev} have been recovered from the coalgebra symmetry
approach.


\section*{Acknowledgements}

{This work was partially supported  by the Ministerio de Educaci\'on y
Ciencia   (Spain, Project FIS2004-07913),  by the Junta de Castilla y
Le\'on   (Spain, Project VA013C05), and by the INFN--CICyT (Italy--Spain).}

\vfill



\begin{thebibliography}{99}
\footnotesize

\bibitem{BHletter}
   A.  Ballesteros and F.J. Herranz, {\it J. Phys.  A:
Math. Theor.} {\bf 40}, F51 (2007).

\bibitem{BHSIGMA}
 O. Ragnisco, A.  Ballesteros, F.J. Herranz and F. Musso, {\it SIGMA} {\bf 3},  026 (2007).


\bibitem{plb} A.  Ballesteros, F.J. Herranz and   O. Ragnisco, {\it Phys. Lett. B} 
{\bf 610}, 107 (2005). 
 
\bibitem{jpa2D}  A.  Ballesteros, F.J. Herranz and   O. Ragnisco,    {\it J. Phys.  A:
Math. Gen.} {\bf 38}, 7129 (2005). 

 \bibitem{Checz}
A.  Ballesteros, F.J. Herranz and   O. Ragnisco,   {\it Czech. J. Phys.}
{\bf 55}, 1327 (2005).

\bibitem{Ohn}
     C.   Ohn,
  {\it Lett. Math. Phys.}  {\bf 25},  85 (1992).



\bibitem{BR}
     A.  Ballesteros and O. Ragnisco,
{\it J. Phys. A: Math. Gen.} {\bf 31},  3791 (1998).

 

\bibitem{Deform}
     A.  Ballesteros and F.J. Herranz,  
{\it J. Phys. A: Math. Gen.} {\bf 32},  8851 (1999).




 \bibitem{darboux1}
E.G. Kalnins, J.M. Kress and P. Winternitz,
{\it J. Math. Phys.} {\bf 43},  970 (2002).

 \bibitem{darboux2}
E.G. Kalnins, J.M. Kress, W. Miller Jr and P. Winternitz,
{\it J. Math. Phys.} {\bf 44}, 5811 (2003).

 \bibitem{darbouxresto}
E.G. Kalnins, J.M. Kress and W. Miller Jr,
{\it J. Math. Phys.} {\bf 46},  053509 (2005); {\it ibid.}
{\it J. Math. Phys.} {\bf 46},  053510 (2005) ;
{\it ibid.}
{\it J. Math. Phys.} {\bf 46},  103507 (2005):
{\it ibid.}
{\it J. Math. Phys.} {\bf 46},  043514 (2006);
{\it ibid.}
{\it J. Math. Phys.} {\bf 46},  093501 (2006).

 \bibitem{fris}
   J.  Fris, V. Mandrosov, Ya A Smorodinsky, M Uhlir and P Winternitz,
{\it   Phys. Lett.} {\bf 16},  354 (1965).

 \bibitem{evans2}
   N.W.  Evans,
{\it   Phys. Lett. A} {\bf 147}, 483 (1990).


\bibitem{enciso}
 A.  Ballesteros, A. Enciso, F.J. Herranz and   O. Ragnisco,  ``{A} novel maximally superintegrable system
in {$n$} dimensions", arXiv:math-ph/0612080.

 \bibitem{ramon} 
F.J. Herranz, R. Ortega and M. Santander, {\it
J.\, Phys.\, A: Math.\, Gen.} {\bf 33}, 4525 (2000).

\bibitem{Conf}    F.J. Herranz and M. Santander,  
{\it
J.\, Phys.\, A: Math.\, Gen.} {\bf 35},  6601 (2002).

 \bibitem{Evansa}
   N.W.  Evans, 
{\it   Phys. Rev. A} {\bf 41},  5666 (1990).

\bibitem{VulpiLett}
 A.  Ballesteros, F.J. Herranz,   M. Santander and T. Sanz-Gil, 
  {\it J. Phys. A: Math. Gen.}  {\bf 36}, L93 (2003). 

\bibitem{CRMVulpi}
 F.J. Herranz,   A.  Ballesteros, M. Santander and T. Sanz-Gil,
{\it Superintegrability in Classical and Quantum   Systems} 
{\it (CRM Proc.~and Lecture Notes  vol.~37)}
ed.~ P. Tempesta {\em et al}
(Providence, RI:   AMS) p.~75 (2004); preprint arXiv:math-ph/0501035.
   
\bibitem{RS}
    M.F.   Ra\~nada and  M. Santander,
{\it J. Math. Phys. } {\bf 40}, 5026 (1999).


 \bibitem{PogosClass1}
E.G. Kalnins, W. Miller Jr   and G.S. Pogosyan,
{\it J. Phys. A: Math. Gen.} {\bf 33},  6791 (2000).


 \bibitem{PogosClass2}
E.G. Kalnins, J.M. Kress, G.S. Pogosyan and W. Miller Jr, 
{\it J. Phys. A: Math. Gen.} {\bf 34},  4705 (2001).

\bibitem{KalninsH2}
E.G. Kalnins, W. Miller Jr   and G.S. Pogosyan,
{\it J. Math. Phys. } {\bf 38},  5416 (1997).

\bibitem{Higgs}
  P.W.   Higgs,
{\it J. Phys. A: Math. Gen.} {\bf 12},  309 (1979).

\bibitem{Leemon}
 H.I.    Leemon,
{\it J. Phys. A: Math. Gen.} {\bf 12},  489 (1979).


\bibitem{Schrodingerdual}
E.G. Kalnins, W. Miller Jr   and G.S. Pogosyan,
{\it J. Math. Phys.} {\bf 41},  2629 (2000). 

 \bibitem{Schrodingerdualb}
A. Nersessian and  G. Pogosyan,
  {\it Phys. Rev. A}  {\bf 63},  020103 (2001).

 
 
\bibitem{kiev} A.  Ballesteros and F.J. Herranz,    {\it
SIGMA} {\bf 2},   010 (2006).







\end{thebibliography}
\end{document}